\documentclass[numbers,sort&compress]{article}
\usepackage[final]{neurips_2024}

\usepackage{pdfpages}
\usepackage[utf8]{inputenc} 
\usepackage[T1]{fontenc}    
\usepackage{hyperref}       
\usepackage{url}            
\usepackage{booktabs}       
\usepackage{amsfonts}       
\usepackage{nicefrac}       
\usepackage{microtype}      
\usepackage{xcolor}
\usepackage{amssymb}
\usepackage{relsize}
\usepackage{amsmath}
\usepackage{bbding}
\usepackage{comment}

\usepackage{tikz}
\usepackage[all]{xy}
\usetikzlibrary{positioning} 
\usetikzlibrary{arrows,decorations.pathmorphing,shapes} 
\usepackage{tikz-cd}

\definecolor{Green}{rgb}{0, 0.6, 0.1}
\definecolor{Orange}{rgb}{1, 0.6, 0}

\usepackage{tikz}

\title{Information in a recurrent Retina-V1 network with \\ realistic noise, feedback and nonlinearities
}

\author{Javier Rodríguez\\
  Image Processing Lab\\
  Universitat de València\\
  Paterna 46980, València, Spain \\
  \texttt{jaroji@alumni.uv.es} \\
  \And
  Raquel Giménez\\
  Image Processing Lab\\
  Universitat de València\\
  \texttt{raquel.gimenez@uv.es} \\
  \And
  Jesús Malo\thanks{Corresponding author. Web Site: \texttt{https://isp.uv.es/excathedra.html}
  } \\
  Image Processing Lab\\
  Universitat de València\\
  \texttt{jesus.malo@uv.es} \\  
  }

\begin{document}
\maketitle

\vspace{-0.4cm}
\begin{abstract}
Quantitative estimation of information flow in early vision with psychophysically realistic networks is still an open issue. This is because, up to date, the necessary elements  (\emph{general and plausible network}, \emph{accurate noise}, and \emph{reliable information measures}) have not been put together. 
As a result, previous works made different approximations that limit the generality of their results.


This work combines the following elements for the first time:
\textbf{(1) \emph{General and plausible recurrent net}}: a cascade of linear+nonlinear psychophysically tuned layers~\cite{Malo06a, Gomez19, Malo20, Li24}, augmented to consider top-down feedback following~\cite{Mobarhan18,Born2021}.
\textbf{(2) \emph{Accurate noise}} in every layer, which is tuned to reproduce psychometric functions both in contrast detection and discrimination following~\cite{Malo25}.
\textbf{(3) \emph{Reliable information measures}} that have been checked with analytical results, both in general~\cite{Laparra24}, and in similar visual neuroscience contexts~\cite{Li24}, and hence can be applied in this (more general) case where analytical results are difficult to obtain.

The joint use of these elements allows a reliable study of information flow depending on different connectivity schemes (different nonlinearities and interactions), different noise sources, and different stimuli. Results show the benefits of feedback in two ways: (1)~the information loss in the data processing inequality along the pathway is reduced by the $\textrm{V1} \rightarrow \textrm{LGN}$ recurrence for values of feedback that give stable steady state solutions, and 
(2)~the stability of the net is assessed through standard stability analysis based on the maximum of eigenvalues and we find an optimal value for the feedback in terms of the accuracy of the reconstructed signal from the cortical representation.


\end{abstract}

\section{Introduction}

\vspace{-0.05cm}
\subsection{Motivation: the ingredients of the information flow problem}

Early vision is about transferring information from the outside world (the input visual stimuli) into the inner signal representation (at certain layer of the visual neural network).
Appropriate quantification of this information transference is crucial in different domains.
On the one hand, it is a fundamental question to characterize information flow between different regions of the visual brain \cite{Seung93,Pouget16,PNAS24}, to measure functional connectivity~\cite{Friston11,Lizier11,QiangEntr22}, to characterize the bottleneck of the system~\cite{Malo20,Malo22}, and to find principled explanations of certain biological behaviors~\cite{Barlow61,Laughlin81,Malo10}.
On the other hand, in image engineering, it is important to quantify the \emph{relevant} (visible) information both in compression (either of images~\cite{Wallace91,Malo06a} or video~\cite{MPEG2,Malo01a}), and in assessing the subjective quality of the results~\cite{Sheikh05,Sheikh06,Malo21}.

However, accurate estimation of the transferred information depends on the consideration of four non-trivial factors: 
\begin{itemize}
     \item \emph{Signal}: modeling the multivariate probability density of the input stimuli.
     \item \emph{Network}: modeling the architecture (transforms of the input signal) including plausible nonlinearities, interactions, and top-down recurrence.
     \item \emph{Noise}: description of the neural noise that explains the variability of the behavioral response.
     \item \emph{Reliable information measures}: if appropriate models of the signals are not available (or are not analytically tractable given the complexities of the network), empirical measures, could be used in large datasets of visual stimuli and the elicited noisy responses.
\end{itemize}

Simultaneous consideration of all the above (multivariate) elements is not easy. Therefore, previous works that addressed the quantification of information flow in early vision made different approximations that limit the generality of their results: 

Specifically, most of the literature on early vision based on information maximization did not consider the noise in the sensors~\cite{Olshausen96,schwartz_natural_2001,Malo06b,Laparra12,Gutmann14,Gomez19}.
The literature that used accurate information measures~\cite{Sheikh05,Sheikh06,Malo20,Malo22,Li24}, did not consider plausible amounts of noise in the network\footnote{For instance, on the one hand, authors in~\cite{Sheikh05,Sheikh06} use Gaussian approximation for the images to get analytical results. Moreover, they claim that the level of Gaussian noise in the wavelet sensors is not relevant as long as it is not zero because it does not have a big impact in reproducing subjective image quality. However, subjective image quality has been questioned as a valid measure of visual behavior~\cite{Martinez19}. 
On the other hand, authors in~\cite{Malo20,Malo22,Li24} assume a standard deviation of $5\%$ of the dynamic range of the signal at each layer, but they recognize they do that  just for convenient illustration, leaving more accurate estimates for further work. }. This also applies to the literature that has considered noisy versions of hierarchical 
models~\cite{Berardino2017,PNAS24}.
Recent works that considered recurrent networks, e.g.~\cite{Gomez19,Malo24}, disregarded noise after the cortical nonlinearities because that complicates the analytical treatment of the problem as seen in~\cite{Li24}: in fact, if noise in the cortical representation and recurrence have to be considered at the same time, (non-plausible) linear models had to be assumed for the sake of analytical tractability. In this way, in~\cite{Li24} two limited classes of models had to be considered: Class I focused on nonlinearities (with no recurrence nor noise after the nonlinearities), and Class II included recurrence and noise but with (too simple) linear assumptions. That work was useful to check that certain information estimates~\cite{Laparra11,Laparra24} are accurate in this context by checking the analytical results of Class I and Class II models with the empirical estimates, but did not apply this empirical tool to more general models, which are not analytically tractable.

In order to overcome the limitations of previous approaches mentioned above, in this work we consider a general, psychophysically-tuned  model that includes both nonlinearities, top-down feedback and psychophysically scaled noise in every layer, and we study the behavior of such model using a reliable empirical estimator of information theoretical magnitudes.  

In particular, we combine:
\begin{enumerate}
\item \emph{Network:} a cascade of linear+nonlinear psychophysically tuned layers~\cite{Malo06a,Gomez19,Malo20,Malo24,Li24}, extended to consider cortico-thalamic feedback~\cite{McClurkin94,Mobarhan18,Born2021,SommerNIPS22}.

\item \emph{Noise:} psychophysically tuned noise to reproduce psychometric functions both in detection and discrimination~\cite{Malo25}.

\item \emph{Reliable information measures}: that have been checked with analytical results~\cite{Laparra24,Li24}.

\end{enumerate}

This (more accurate and general) setting allows the exploration of the effects of different biologically-relevant factors on the transmitted information, namely: 
(a)~bandwidth of the LGN bottleneck (size of center-surround cells or Contrast Sensitivity Function).
(b)~Nonlinearity in the retinal front-end.
(c)~Nonlinearity in the cortical sensors and their interaction as described in Divisive Normalization or Wilson-Cowan models.
(d)~Role of simple top-down feedback schemes.
(e)~Role of different noise models: Gaussian and Poisson variability including correlation in the noise patterns.
(f)~Data processing inequality in early vision.
(g)~Information transmitted for stimuli of different nature.

\subsection{Structure of the paper}

Section~2 first describes the retina-LGN-V1 model that includes Divisive Normalization nonlinearities extended with recurrent top-down feedback from V1. This model unifies and generalizes the (limited) model classes considered in [Neurocomp.24]. 
Then, the appropriate noise level in each layer of the model is fitted to reproduce behavioral uncertainties following the procedure proposed in [JoV.25].
Section~3 reviews the empirical mutual information measure [IEEE PAMI 24] to be used in the experiments.
Section~4 shows the results of the stability analysis, optimal feedback and preliminary plots on the data processing inequality in the pathway. Finally, 
Section~5 discusses the implications of the tools presented here on an information-theory-based research agenda of this system: the partition of the stimulus space, the range of parameters considered for the LGN, V1, feedback and noise schemes.

\section{The model: noisy, nonlinear, recurrent retina-V1 pathway}

\subsection{Physiologically-Psychophysically sensible transforms}

Consider the standard feedforward retina-V1 pathway~\cite{Martinez18,Martinez19} extended to include known feedback from V1~\cite{McClurkin94,Mobarhan18,Born2021,SommerNIPS22}, generically as considered in~\cite{Li24}:

\begin{equation}\label{}
\xymatrix{\mathrm{Retina} \ar[r] &  \mathrm{LGN} \ar[r]  & \ar@/^1.25pc/[l] 
\mathrm{V1}}
\end{equation}

\begin{equation}\label{Framework}
\xymatrix{\mathbf{a} \ar[r] & \mathbf{b} \ar[r]^{c_{2}} &  \mathbf{c} \ar[r]^{c_{3}}  & \mathbf{d} \ar[r]^{c_{4}, c_H}  & \ar@/^1.5pc/[lll]^{c_{1}} 
\mathbf{e}}
\end{equation}

However, we unify and generalize Models I and II in~\cite{Li24}, by considering nonlinearities, recurrence and noise in every layer at the same time:

\begin{align}\label{eq:model}
    \mathbf{b}(t) &= \mathbf{a}(t)^{\gamma_b} + \mathbf{n_b}(t) + \frac{c_{1}}{c_{2}c_{3}c_{4}} F^{-1} \cdot \mathbf{e}(t-\Delta t), \nonumber \\
    \mathbf{c}(t) &= c_{2} K \cdot \mathbf{b}(t) + \mathbf{n_c}(t) = c_{2} F^{-1} \cdot \lambda_{\text{CSF}} \cdot F \cdot \mathbf{b}(t) + \mathbf{n_c}(t), \\[0.2cm]
    \mathbf{d}(t) &= c_{3} F \cdot \mathbf{c}(t) + \mathbf{n_d}(t), \nonumber \\
    \mathbf{e}(t) &= c_{4} \mathbb{K}(\mathbf{d^\star} \cdot \operatorname{sign}(\mathbf{d}(t)) \cdot \frac{|\mathbf{d}(t)|^{\gamma_d}}{\beta + c_{H} H \cdot |\mathbf{d}(t)|^{\gamma_d}} + \mathbf{n_e}(t)= c_4 \mathcal{N}(\mathbf{d}(t))+ \mathbf{n_e}(t). \nonumber
\end{align}

Model parameters (except for the feedback connection and the noise levels) were justified in~\cite{Li24} through the reproduction of human psychophysics on image distortion.

The following sections justify the remaining parameters: feedback and noise levels at the retina and the inner representation.

\subsection{Physiological-mathematical grounds of the feedback model}

The feedback in our model (originally from~\cite{Li24}) inverts the local-frequency signal representation of V1 before injecting the signal back at the input of LGN.
This was done just because of mathematical convenience: as the signal at the LGN is represented in the spatial domain, one cannot expect a direct injection of the V1 signal since the latter is in a different representation.

\begin{figure}[t]
\begin{center}
\vspace{-0.3cm}
\includegraphics[width=1\textwidth]{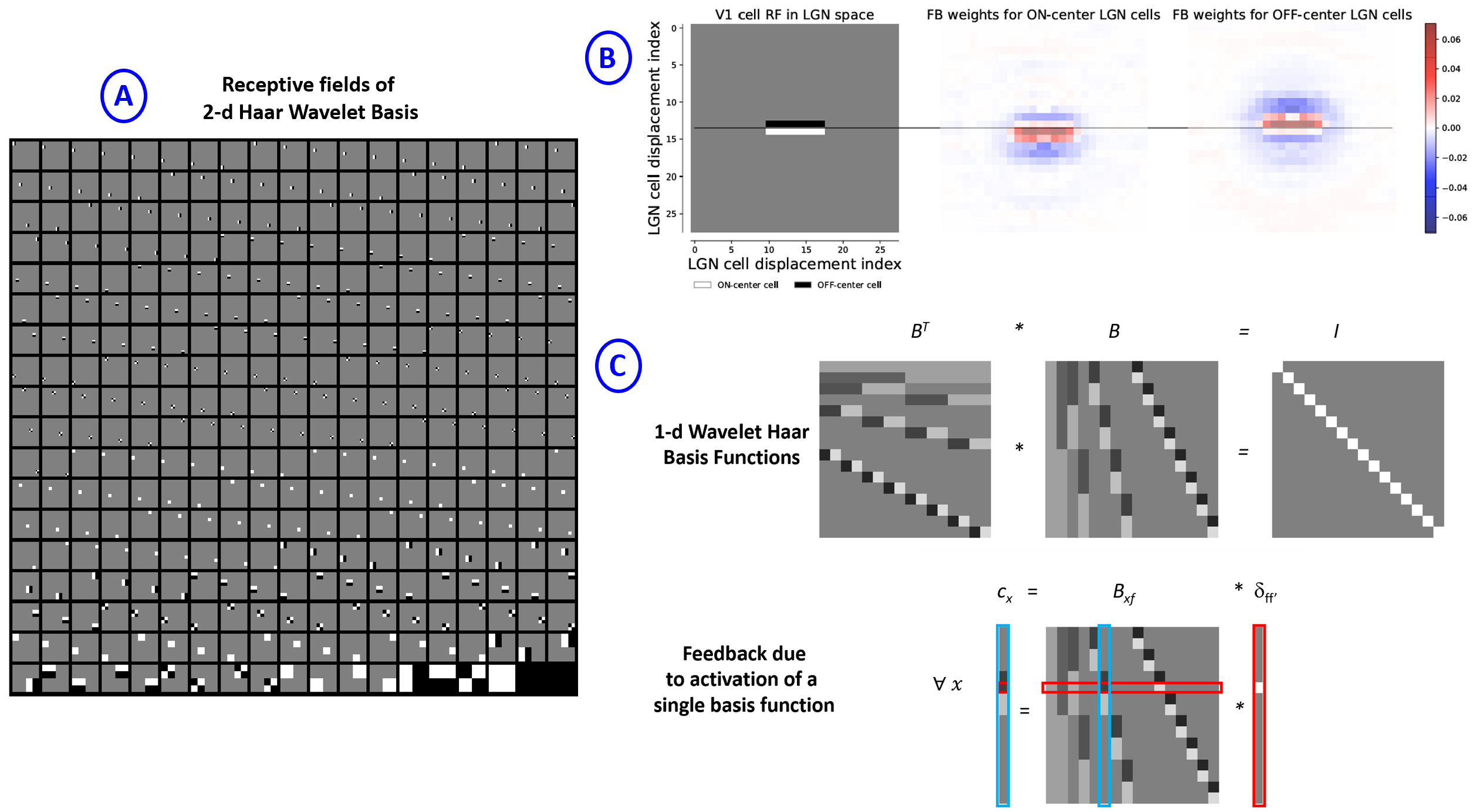}
\vspace{-0.3cm}
\caption{\small{\textbf{The feedback learned in~\cite{SommerNIPS22} can be seen as an inverse transform}. (a)~Illustrative set of V1-like connectivity weights in $LGN \rightarrow V1$: a local-frequency transform. In this illustration, an orthonormal set of Haar functions compatible with the model used in~\cite{SommerNIPS22}. (b)~Result of~\cite{SommerNIPS22}: learned weights in the $V1 \rightarrow LGN$: the influence of the response of a V1 neuron in LGN strongly resembles the receptive field of the V1 neuron. (c)~One dimensional example of the Haar transform showing that such feedback weights are actually implementing an inverse transform.}}
\label{feedback_inverse}
\vspace{-0.4cm}
\end{center}
\end{figure}

This "\emph{inverse feedback circuit}" purely based on a domain-consistency argument actually has physiological and statistical grounds. 
Fig.~\ref{feedback_inverse}.a shows illustrative weights of the forward, $LGN \rightarrow V1$, transform.
Fig.~\ref{feedback_inverse}.b, shows the statistically learned weights of the feedback $LGN \leftarrow V1$ in~\cite{SommerNIPS22}, which are consistent with physiology~\cite{Born2021}.
Interestingly, Fig.~\ref{feedback_inverse}.c shows that in case of V1-like forward receptive fields, such backward connectivity and influence can be interpreted as implementing the inverse of the forward transform. 

Therefore, the negative/positive regions in the V1 backward connections inhibit/excite the input of LGN, and this is what the orthogonal inverse, $F^{-1} = F^\top$, does. Therefore, the feedback using the inverse proposed in~\cite{Li24} makes sense (not only dimensionally), but also physiologically.
Incidentally, the appendix makes some considerations on how the feedback could adaptively modify the size of the center-surround receptive fields of the LGN in line with the discussions in~\cite{SommerNIPS22,Born2021}. 

\subsection{Psychophysical tuning of the noise in each layer} 

We recently proposed a method to estimate the early (retinal) and the late (cortical) noise sources which are compatible with accurate psychophysics of contrast detection and discrimination~\cite{Malo25}.

\begin{figure}[t]
\begin{center}
\vspace{-0.3cm}
\includegraphics[width=1\textwidth]{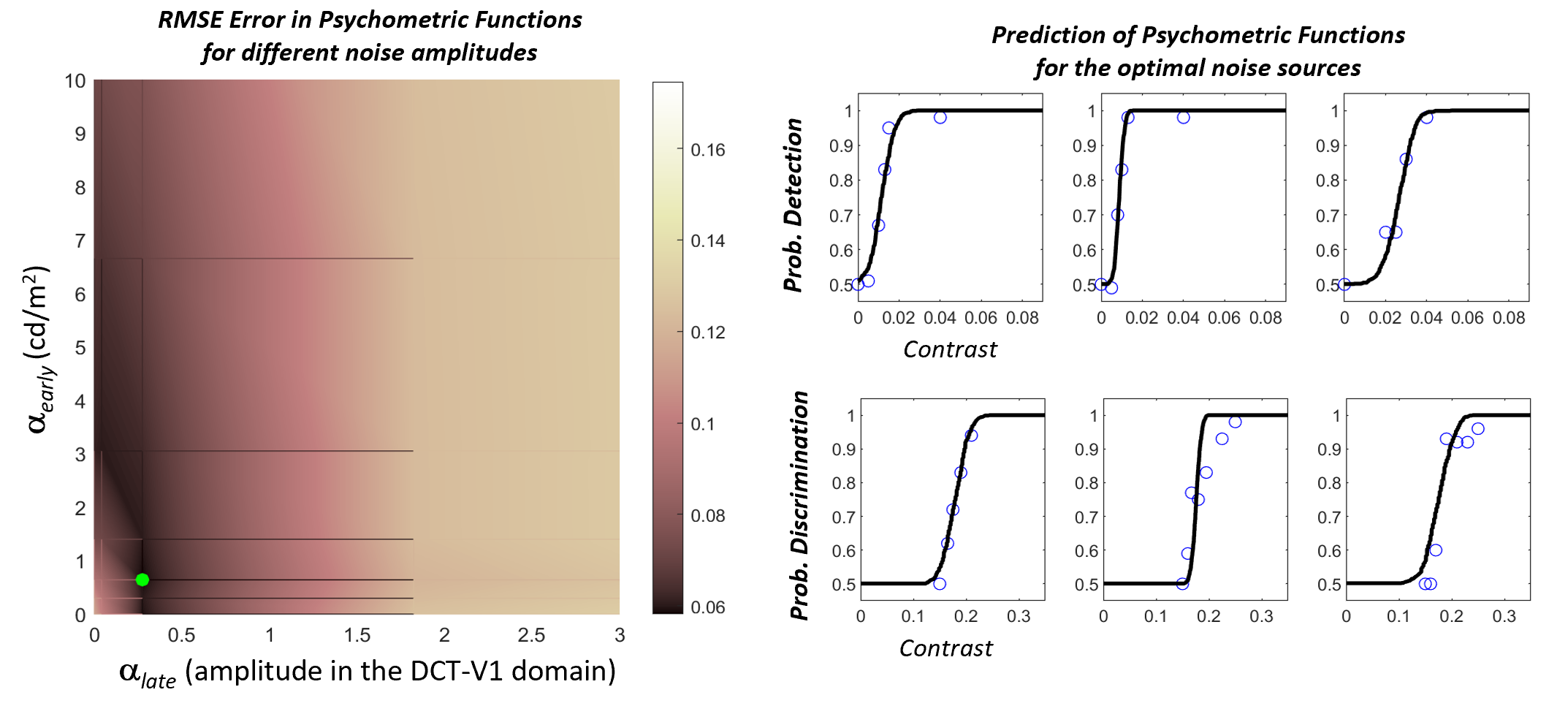}
\vspace{-0.3cm}
\caption{\small{\textbf{Estimation of early and late noise in our models}. \emph{Left panel} highlights in green the point indicating the optimal amplitudes of the early and late noise sources (vertical and horizontal axis respectively) which are compatible with the psychophysical behavior (lower error is better). \emph{Right panel} shows the corresponding fit (black curves) of the experimental psychometric functions of~\cite{Wichmann02} (blue circles) as prescribed in~\cite{Malo25}.}}
\label{noise}
\vspace{-0.4cm}
\end{center}
\end{figure}

We applied that method for the network considered in this paper.
Fig.~\ref{noise} (left) highlights the amplitudes of the early and late noise sources (vertical and horizontal axis respectively) and Fig.~\ref{noise} (right) shows the corresponding fit of the psychometric functions of~\cite{Wichmann02} as prescribed in~\cite{Malo25}. 

In the simulations below the uncertainty in each spatial or local-frequency layer has been set according to these estimates.

\section{Methods: transmitted information from multivariate Gaussianization}

The quantification of the information in the input signals, $\mathbf{s}$, and its degradation along the network, $\mathbf{x}$~(retinal ganglion cells), $\mathbf{y}$~(LGN), and finally, $\mathbf{e}$ and $\mathbf{z}$ (linear and nonlinear V1, respectively), requires estimating info-theoretic variables.
On the one hand, the \emph{entropy} of the input stimuli, $H(\mathbf{a})$, and on the other hand, the information shared between the stimuli (the outside world) and the signal at different layers along the pathway (the inner representation of the world), i.e. the \emph{mutual information} values $I(\mathbf{a},\mathbf{b})$, $I(\mathbf{a},\mathbf{c})$, $I(\mathbf{a},\mathbf{d})$, and $I(\mathbf{a},\mathbf{e})$.  

As opposed to the analytical approach in~\cite{Li24}, here we take a purely numerical estimation based on image samples and the corresponding responses as in~\cite{Malo20}. 
We do so in this case because here 
we unify the model classes that were considered separately in~\cite{Li24} for analytical convenience. The more general and accurate recurrent-and-nonlinear model considered here complicates the analytical approach.

Our numerical method to estimate entropy, $H$, and transmitted information, $I$ is based on the fact that both (hard to compute) \emph{multivariate} quantities can be put in terms of (easy to compute) \emph{univariate} quantities and \emph{Total Correlation}, $T$, which is easy to compute using a specific  Gaussianization, $\mathcal{G}$.

Lets review first how a specific Gaussianization (our Rotation-Based Iterative Gaussianization, RBIG~\cite{Laparra11}) is actually better than other differentiable Gaussianization transforms to estimate $T$. Then, we follow~\cite{Laparra24} to recall the expressions of the quantities that we require here, $H$, and $I$, in terms of the quantity that we can easily compute, i.e. $T$.

\subsection{Total correlation from Gaussianization transforms and from RBIG}

A Gaussianization adapted to certain  $d$-dimensional random variable $\mathbf{v}$ coming from any PDF is a mapping, $\mathcal{G}_\mathbf{v}$. that transforms samples of this variable into $d$-dimensional samples $\mathbf{g}$ coming from a unit-covariance Gaussian:
\begin{equation}
\mathbf{v}\xrightarrow{\,\,\,\,\mathcal{G}_\mathbf{v}\,\,\,\,}\mathbf{g}_\mathbf{v}
\end{equation}

Gaussianization transforms may be useful to compute $T$ because they transform the input, for which redundancy may be high, $T(\mathbf{v}) \gg 0$, into a vector of independent components, with $T(\mathbf{g}_\mathbf{v}) = 0$. Variation of $T$ under differentiable transforms implies that~\cite{Laparra24}:
\begin{equation}
    T(\mathbf{v}) = \sum_{i=1}^{d} H(v_i) - \frac{d}{2} \log(2\pi e) + \mathbb{E}_\mathbf{v} \Big( \log |\nabla \mathcal{G}_\mathbf{v}(\mathbf{v})| \Big) \label{eq:T_definition}
    \nonumber
\end{equation}

which, in general is not easy to compute because while the marginal (univariate) entropies $H(v_i)$ are easy to compute, the expectation over the whole space, $\mathbb{E}_\mathbf{v} \Big( \log |\nabla \mathcal{G}_\mathbf{v}(\mathbf{v})| \Big)$, definitely is not. 
Fortunately, RBIG is a specific $\mathcal{G}_\mathbf{v}$ which breaks the multivariate Gaussianization in $n$-steps made of simple univariate Gaussianizations plus a rotation, and hence the above non-feasible expression reduces to a sum of variations, $\Delta T^{(j)}$, in each step, $j$, which reduces to easy (univariate) entropy estimations~\cite{Laparra24}: 
\begin{equation}
            T(\mathbf{v}) = \sum_{j=0}^{n-1} \Delta T^{(j)} = \frac{(n-1) d}{2} \log(2\pi e) - \sum_{j=1}^{n} \sum_{i=1}^{d} H(v_i^{(j)})
            \label{TRBIG}
\end{equation}
where $v_i^{(j)}$ with $i=1,\ldots,d$, are the components of the vector at step $j$ in its transit from $\mathbf{v}$ to $\mathbf{g_v}$.

This is the \emph{multivariate-to-univariate} property of RBIG that enables robust estimations of mutivariate information-theoretic quantities otherwise hard to compute.

\subsection{\emph{H} and \emph{I} in terms of \emph{T} for their computation using RBIG}

Basic definition of info-theoretic magnitudes imply a relation between the joint entropy (multivariate) and the total correlation together with marginal entropies~\cite{Laparra24}:
\begin{equation}
   H(\mathbf{v}) = \sum_{i=1}^{d} H(v_i) - T(\mathbf{v}).
   \label{entropyRBIG}
\end{equation}
And, as $T$ can be computed with RBIG using Eq.~\ref{TRBIG}, the (in principle hard to compute) joint entropy reduces to one Gaussianization and univariate entropy estimations.

According to the Proposition in Eq.~11  of~\cite{Laparra24}, 
the mutual information between two variables is equivalent to the 
the total correlation remaining in the vector formed by stacking the two independently Gaussianized variables, i.e.:
\begin{equation}
I(\mathbf{u}, \mathbf{v}) = T([\mathbf{g_u},\mathbf{g_v}]),
\label{I_rbig}
\end{equation}
where $\mathbf{g_u} = \mathcal{G}_\mathbf{u}(\mathbf{u})$ and $\mathbf{g_v} = \mathcal{G}_\mathbf{v}(\mathbf{v})$. Therefore, the computation of $I(\mathbf{u}, \mathbf{v})$ is reduced to a single estimation of $T$ of the stacked Gaussianized variables. Which, if done using RBIG, according to Eq.~\ref{TRBIG}, it reduces to a set of easy (univariate) operations.
As a result, the estimation of $I(\mathbf{u},\mathbf{v})$ implies three Gaussianizations: one to get $\mathbf{g_u}$, another to get $\mathbf{g_v}$, and a final one to compute $T([\mathbf{g_u},\mathbf{g_v}])$.

\section{Experiments and Results}

\subsection{Stability of the network  and optimal feedback}

Recurrent nonlinear networks may be unstable depending on the properties of the feedback. In this section (1) we derive the analytical stability condition for the considered network, (2) we numerically check the correctness and consequences of such condition, and (3) we study the quality of the reconstructed signal depending on the feedback strength. 

In order to apply the standard stability analysis for dynamical systems~\cite{Hubbard1995} we have to reformulate Eqs.~\ref{Framework} as a differential equation. By considering the temporal evolution of the output given the recurrence, it is easy to see that for low-energy signals (i.e. when  the divisive normalization can be linearized as $\mathcal{N}(\mathbf{e}) \approx \nabla_\mathbf{e} \, \mathcal{N}(\mathbf{e}) \cdot \mathbf{e}$), we have the following equation:
\begin{equation}
      \dot{\mathbf{e}} = \mathcal{C} + \left( c_1 \nabla_e \, \mathcal{N}\cdot\lambda_{\mathrm{CSF}} - \mathcal{I} \right)\cdot\mathbf{e} 
      \nonumber
\end{equation}
Therefore, the factor that determines the stability, i.e. the Jacobian of the evolution equation with the signal, here depends on the strength of the feedback, $c_1$, and the slope of the Divisive Normalization and the strength of the CSF bottleneck:
\begin{equation}
      J = c_1 \nabla_e \, \mathcal{N}\cdot\lambda_{\mathrm{CSF}} - \mathcal{I}
      \label{condition}
\end{equation}

Some intuitive conclusions may be extracted from Eq.~\ref{condition}.
In order to be in the stable region, the maximum eigenvalue should be negative, ~\cite{Hubbard1995}. Our Jacobian initially tends to a stable point, $\nabla_{\mathbf{e}} G(\bar{\mathbf{e}}) \approx -I$, when the feedback strength is completely suppressed ($c_1 \to 0$). In this context, we obtain $Re(\lambda_i) = -1$ for every $i$, ensuring absolute stability. However, the exact boundary of unstability ($Re(\lambda_i) = 0$) is dynamically dictated by the local stimulus and the noise injected into the network.

\begin{figure}[htbp]
    \centering
    \begin{minipage}{0.48\textwidth}
        \centering
        \includegraphics[width=\linewidth]{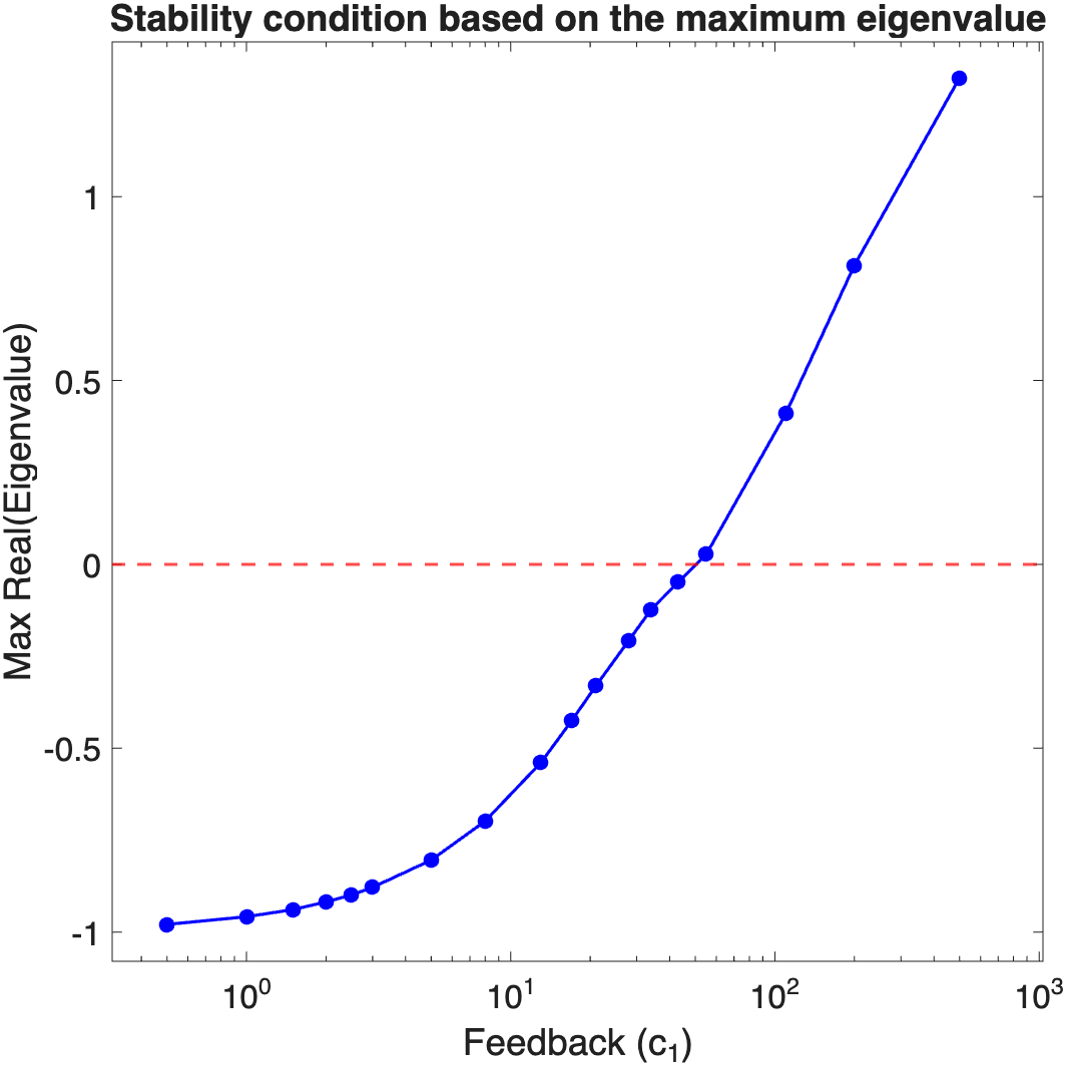}
    \end{minipage}
    \hfill 
    \begin{minipage}{0.48\textwidth}
        \centering
        \includegraphics[width=\linewidth]{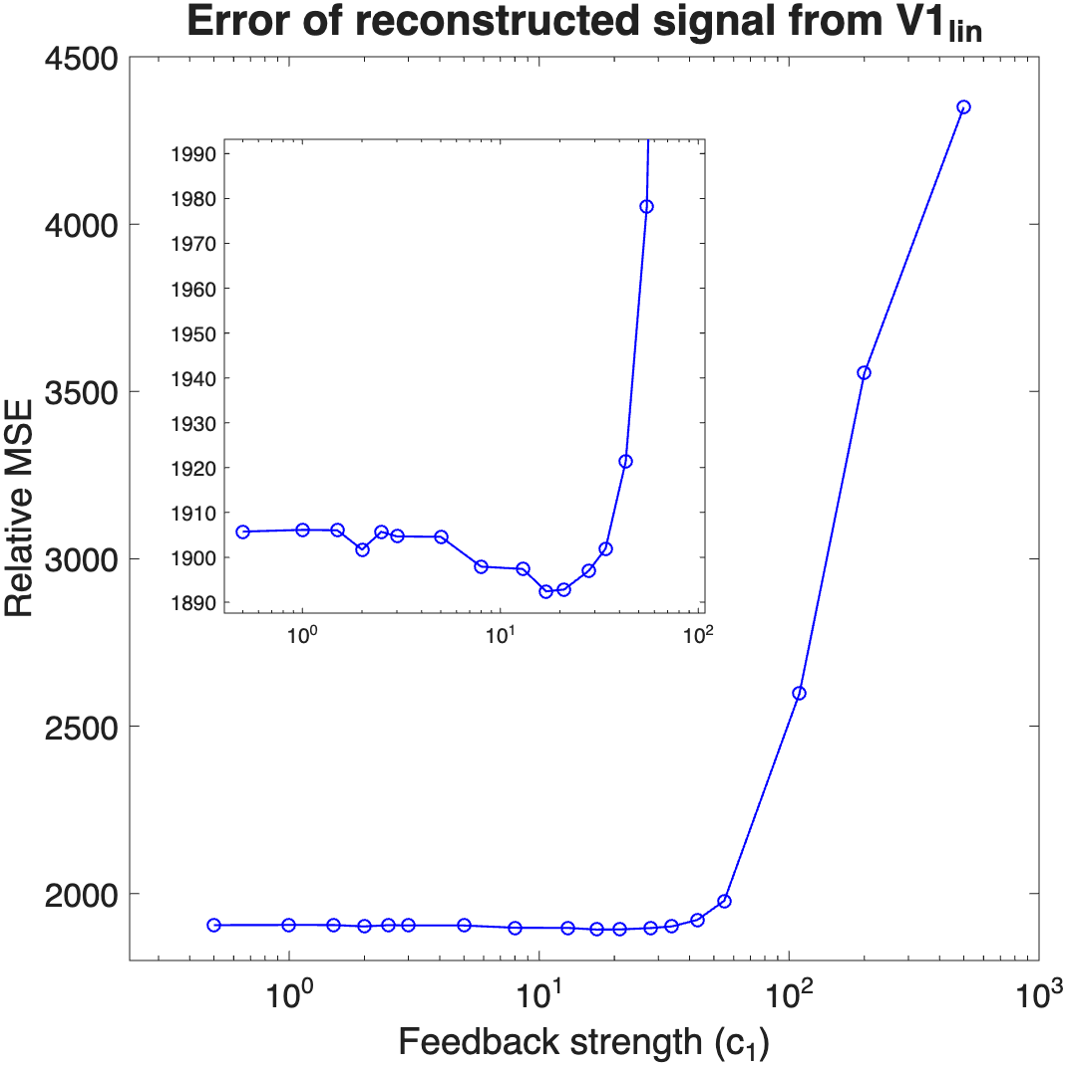}
    \end{minipage}
    
    \caption{\small{\textbf{Stability and reconstruction error depending on feedback}. \emph{Left} Stability Diagram for different feedback strengths in which the dotted red line marks the boundary between stable ($Re(\lambda_i)<0$) and unstable ($Re(\lambda_i)>0$) regime. \emph{Right} Error if we reconstruct the signal from the noisy V1 linear layer for different feedback strengths including the no-feedback $c_1 = 0$ scenario. Quality of the reconstructions has an optimum for $c_1>0$, which differs from the no-feedback, $c_1 = 0$, scenario.}}
    \label{stab}
\end{figure}

This intuition is confirmed by the experimental results shown in Fig.~\ref{stab}~(left) computed from a set of images and a number of feedback strengths. On top of the intuitive push to the right, larger feedbacks also lead to higher eigenvalues. Thus making the reconstruction error larger for higher eigenvalues.

In realistic noisy scenarios, such as the one considered here, being too close to the limit may have catastrophic consequences as empirically shown by reconstructing the signal from an internal layer above the feedback loop. 

In our case we illustrate that by taking the noisy representation, $\mathbf{e}$, after a large enough number of feedback loops and inverting both the local-frequency transform and the CSF filter.
Then, the difference between the reconstructed signal and the actual inuput is a measure of the consequences of the feedback, either benefits or problems.
This difference (Relative MSE) is represented in Fig.~\ref{stab}~(right) for different feedback strengths.

While small feedback strengths keep the quality of the reconstructed signal, the error explodes for too-high feedback strengths, as expected from the approximation to the unstability limit in Fig.~\ref{stab}~(left).
Interestingly, certain values of feedback lead to better reconstruction that the no-feedback scenario.

\begin{figure}[b]
\begin{center}
\vspace{-0.3cm}
\includegraphics[width=1\textwidth]{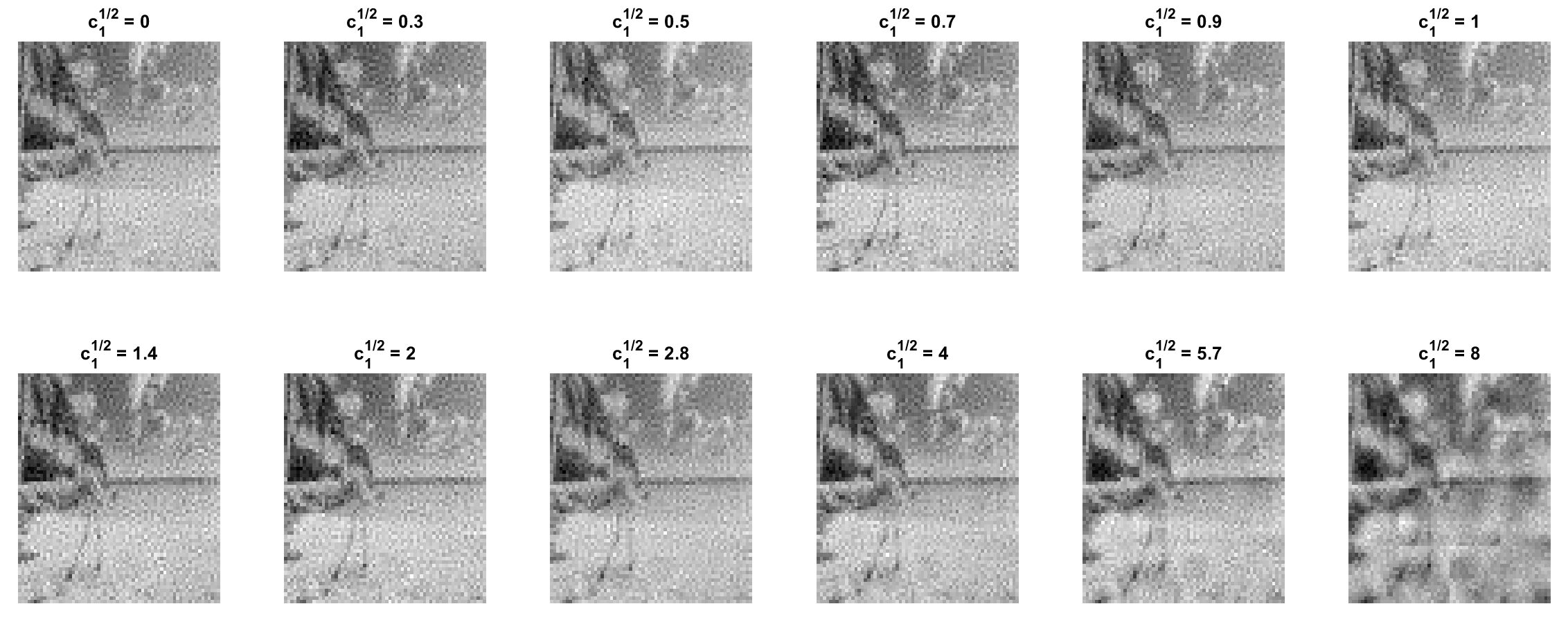}
\vspace{-0.3cm}
\caption{\small{\textbf{Illustrative reconstructed images assuming different feedback strengths}. Quality of the reconstructions has an optimum for $c_1>0$, which differs from the no-feedback, $c_1 = 0$, scenario.}}
\label{stab2}
\vspace{-0.4cm}
\end{center}
\end{figure}

Noise amplification, unstability and error explosion for too-big feedback strengths can be visualized in the reconstructions shown
in Fig.~\ref{stab2}.

\subsection{Data Processing Inequality in the retina-LGN-V1 path}

Here we apply the concepts recalled in Section~3 to study the information flow along the considered network.

Following the empirical checks in~\cite{Li24}, analysis is based on how $5 \cdot 10^5$ image samples of $d = 256$ ($16\times16$ blocks) go through the network considered here. Then, RBIG is applied to compute the information measures as described in Section~3.

\begin{figure}[t]
\begin{center}
\vspace{-0.3cm}
\includegraphics[width=0.7\textwidth]{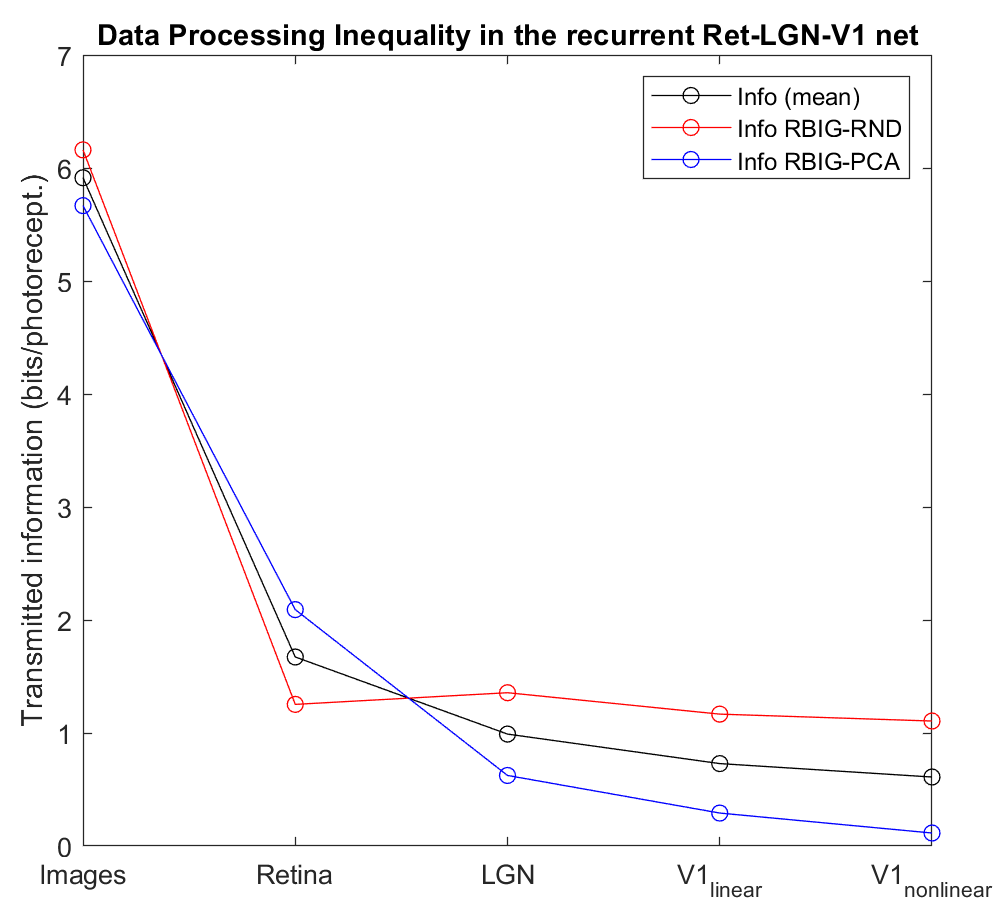}
\vspace{-0.3cm}
\caption{\small{\textbf{Verification of the Data Processing Inequality}. Testing the data processing inequality by estimating Transmitted information in the Retina - V1 path using random rotations (red curve) or the principal component rotations (blue curve) for RBIG.}}
\label{flow}
\vspace{-0.1cm}
\end{center}
\end{figure}

The result of such computations for the system with optimal feedback strength is shown in Fig.~\ref{flow}. This plot shows a behavior which is consistent with the Data Processing Inequality~\cite{Cover06}.

\section{Discussion and conclusions}

In this work we present a scenario for the study of the information flow in the retina-LGN-V1 path considering together the following elements for the first time: (1)~a model with realistic feedback, (2)~realistic noise levels, and (3)~an accurate information estimation technique.

The results of the stability, reconstruction and information experiments suggest a series of conclusions and the proposed scenario and tools open a research agenda for further work.

The optimal value for the feedback strength is about $c_1 = 17$ as seen in Figure~\ref{stab} (\textit{right}). This makes qualitative sense because of the following. The noisy signal in the spatial representation (at LGN) is combined with feedback coming from cortical areas after inversion of the local-frequency transform. The feedback signal is also noisy, but the value $c_1 = 17$ implies that the scale of the feedback signal is bigger to the incoming (forward) signal, and this implies that while part of the signal depending on the real stimulus is reinforced, sum of different realizations of the noise at different neural layers will probably help to reduce it. This explains that the optimal reconstruction error is obtained for non-zero feedback.

On the other hand, a too strong feedback pushes the system towards unstable regions of the differential equation, thus being undesirable as well. All these reasonable denoising benefits of moderate feedback are consistent with the distinction between physiological and psychophysical levels of noise~\cite{Malo25}, and consistent with an eventual denoising goal of the LGN bottleneck~\cite{Li92,Li22}.

Moreover, these results on noise cancellation due to the feedback are a different complementary evidence of the benefits of feedback beyond the change of the size of the center-surround receptive fields~\cite{Born2021,SommerNIPS22}.

These sensible results, and the reasonable evolution of the information flow along the neural path suggest a series of experiments to exploit the presented scenario and tool.

(a)~The experiments presented here could be done for higher (physiologically plausible) noise levels~\cite{Malo25} for a better check of the denoising goal. The optimum from the error function should be stronger for higher noise levels. 

(b)~Stability and reconstruction error experiments could be done for different parameters of the divisive normalization. If the current parameters (optimal for subjective image quality) are modified, what is the impact in the stability of the system?

(c)~Information flow and error could be studied for different number of feedback loops (or feedback time). Maybe there is a relation between strength of feedback and number of feedback loops. Are the reconstruction error benefits visible using the information measures with different feedback strengths? 

(d)~The optimal bandwidth of the LGN bottleneck (size of center-surround cells or Contrast Sensitivity Function) could be studied.

{\color{black} \emph{CSF bandwidth is optimal in information-theoretic terms}: lower cut-off frequencies lead to information loss (signal is discarded), and higher cut-off frequencies lead to information loss as well: input noise propagates ruining the inference.
This would resonate with~\cite{Li92, Li22}, but using quantitative measures of information and proper nolinear models and noise.}

(e)~Nonlinearity in the retinal front-end.

{\color{black} \emph{Weber saturation is optimal in information-theoretic terms} to deal with high-dynamic range stimuli (maybe we have to select stimuli to point that out: e.g. scenes with structures in shadow.). This would connect with~\cite{Laughlin81,Hernandez23,Hernandez24} but using information theoretic measures and a more general model with proper noise.}

(f)~Nonlinearity in the cortical sensors and their interaction as described in Divisive Normalization or Wilson-Cowan models.

{\color{black} \emph{Cortical nonlinearities are optimal in information-theoretic terms} e.g. necessary to deal with low-contrast scenes (maybe we have to select stimuli to point that out: e.g. scenes in fog/rain.). This would connect with~\cite{Hernandez23,Hernandez24} but using information theoretic measures and a more general model with proper noise.}

{\color{black} \emph{Cortical interaction lengths and the strength of the nonlinearity is optimal in info-theoretic terms}. This would connect with the correlation measures checked in \cite{Malo20, Li24} using proper info measures, model and noise.}

(g)~The role of simple top-down feedback schemes.

{\color{black} \emph{V1 -> LGN feedback is necessary to increase transmitted information}. Sharpness of the LGN receptive fields. Maybe we need to increase the feedback a bit to see this.
Note that expressions already show that the feedback implies high-pass after center surround receptive fields!
We could confirm this with information measures and we will be improving the NIPS 22 results!}

(h)~Role of different noise models: Gaussian and Poisson variabilities including correlation in the noise patterns.

{\color{black} On top of independent noise in cortical sensors we could play with correlation related to frequency as in the kernel of Divisive Normalization. The sexy result would be: \emph{Noise correlations in the cortex maximize transmitted information}}

(i)~Data processing inequality in early vision.

{\color{black} \emph{Proper quantification of information loss} (as opposed to~\cite{Malo20} -in the good direction but limited-, and completely improving Zhong et al. Cogn. Neurodyn. 21 -which is totally wrong!-).}

(j)~Information transmitted for stimuli of different nature.

{\color{black} \emph{Transmitted information in early is bigger for more probable stimuli} as suggested in ~\cite{Martinez18, Malo20, Laparra24}, but now in a better context.}

\enlargethispage{20pt}

\ack{The work was partially funded by the Spanish Government and the EU under the  MCIN/AEI/FEDER/UE Grant PID2023-152133NB-I00,  and by the BBVA Foundations of Science program in Maths, Stats, Comp. Sci. and AI, grant VIS4NN: \emph{Vision Science for Artificial Neural Networks}.}

\bibliographystyle{unsrt}
\bibliography{actual5,biblio_clean,early_late_references,referencesnodoi}


\clearpage

\appendix

\section{Appendix: LGN receptive fields could shrink due to feedback from V1}

Remember the receptive field of neurons at certain layer is the linear approximation of the response (wrt the input)~\cite{Ringach02,Martinez18}. In our model with recurrence the LGN is layer $y$, then:

\begin{eqnarray}
\frac{\partial \mathbf{y}(t)}{\partial \mathbf{s}(t)} & = & \frac{\partial \mathbf{y}(t)}{\partial \mathbf{x}(t)} \cdot \frac{\partial \mathbf{x}(t)}{\partial \mathbf{s}(t)} = c_{xy} \, K \cdot \left(\gamma_x \mathbb{D}_{s^{\gamma_x-1}} + \frac{c_{zx}}{c_{xy}\,c_{ye}\,c_{ez}} \, F^{-1} \cdot \frac{\partial \mathbf{z}(t-\Delta t)}{\partial \mathbf{s}(t)} \right) \nonumber \\
 & = & c_{xy} \, K \cdot \left(\gamma_x \mathbb{D}_{s^{\gamma_x-1}} + \frac{c_{zx}}{c_{xy}\,c_{ye}\,c_{ez}} \, F^{-1} \cdot \frac{\partial \mathbf{z}}{\partial \mathbf{e}} \cdot \frac{\partial \mathbf{e}}{\partial \mathbf{y}} \cdot \frac{\partial \mathbf{y}}{\partial \mathbf{x}} \cdot \frac{\partial \mathbf{x}}{\partial \mathbf{s}}\right) \nonumber \\
  & = & c_{xy} \, K \cdot \left(\gamma_x \mathbb{D}_{s^{\gamma_x-1}} + \frac{c_{zx}}{c_{ez}} \, \, F^{-1} \cdot \frac{\partial \mathbf{z}}{\partial \mathbf{e}} \cdot F \cdot K \cdot \frac{\partial \mathbf{x}(t-\Delta t)}{\partial \mathbf{s}(t)}\right) \nonumber \\
  & = & c_{xy} \, K \cdot \left(\gamma_x \mathbb{D}_{s^{\gamma_x-1}} + \frac{c_{zx}}{c_{ez}} \, \, F^{-1} \cdot \frac{\partial \mathbf{z}}{\partial \mathbf{e}} \cdot F \cdot K \cdot \gamma_x  \mathbb{D}_{s^{\gamma_x-1}}\right) \nonumber \\
  & = & c_{xy} \, K \cdot \left( I + \frac{c_{zx}}{c_{ez}} \, \underbrace{F^{-1} \cdot \frac{\partial \mathbf{z}}{\partial \mathbf{e}} \cdot F}_{\Gamma = high-pass?} \cdot K \right) \cdot \gamma_x \mathbb{D}_{s^{\gamma_x-1}}
\label{RF}
\end{eqnarray}

where we made two assumptions for a cleaner result with simpler interpretation: (i) the noises are independent of the signals, and (ii) 2-back inner responses $z(t-2\Delta t)$ are decoupled from $s(t)$. The second assumption should be true for large enough temporal delays, but here we reduced this relation to a single step in discrete time, $\Delta t$, to avoid further recurrence in the expression and hence simpler interpretation. If the decoupling assumption is relaxed and we consider relations at previous discrete times one would have a recurrent expression like: 

\begin{eqnarray}
\frac{\partial \mathbf{y}}{\partial \mathbf{s}} & = &  c_{xy} \, K \cdot \left( I + \frac{c_{zx}}{c_{ez}} \, \Gamma  \cdot K \left( I + \frac{c_{zx}}{c_{ez}} \, \Gamma  \cdot K \left( I + \frac{c_{zx}}{c_{ez}} \, \Gamma  \cdot K\right) \right) \right) \cdot \gamma_x \mathbb{D}_{s^{\gamma_x-1}}
\label{RF2}
\end{eqnarray}

where the kernel $K$ is modified by many (instead of one) high-pass versions of the original $K$. Note that in order to obtain shrpener versions of the center-surround receptive fields $\Gamma = F^{-1} \cdot \frac{\partial \mathbf{z}}{\partial \mathbf{e}} \cdot F$ should be  a high-pass filter. Some results~\cite{Martinez18} suggest that this could be the case in larger scale models, but is it is not that obvious in this small scale model for $16\times16$ blocks.

Eqs.~\ref{RF} and~\ref{RF2} mean that the wired center-surround connections in the retina-LGN transform, which are in the (band-pass) kernel K are not the only element of the LGN receptive fields: given the nature of the cortical nonlinearity, there is an extra term in which a high-pass filter is applied to $K$. Note that the final matrix $\gamma_x \mathbb{D}_{s^{\gamma_x-1}}$ is just a diagonal matrix that describes changes in the gain of the receptive fields in regions of different luminance (Weber law), but it does not depend on the feedback. 
This second term in the parenthesis adds a high-pass contribution to the band-pass shape of the kernel K, thus reducing the size of the receptive fields as reported in the experimental literature~\cite{Born2021,SommerNIPS22}.

Note that signal-dependent noise sources (e.g. Poisson), or considering many terms in the recurrence would not substantially change the high-pass modification of $K$.

\end{document}